\documentclass[parskip=no]{article}

\usepackage{fzjdoc}
\usepackage{cases}
\usepackage{float}
\usepackage{hyphenat}
\begin{document}

%
%
%
%
%
%
%
%
%
%
%
%









\newpage

\begin{center}
    \fontsize{18pt}{18pt}\selectfont
    \textbf{Interplay of Crystallization and Amorphous Spinodal Decomposition during Thermal Annealing of Organic Photoactive Layers}
    \par
    \vspace{0.3cm}
    \fontsize{12pt}{12pt}\selectfont
    {Maxime Siber,$^{\ast}$\textit{$^{a,b}$} Olivier J. J. Ronsin,\textit{$^{a}$} Gitti L. Frey,\textit{$^{c}$} and Jens Harting \textit{$^{a,b,d}$}}
\end{center}

\fontsize{9pt}{9pt}\selectfont
{\textit{$^{a}$~Helmholtz Institute Erlangen-Nürnberg for Renewable Energy, Forschungszentrum Jülich, Fürther Straße 248, 90429 Nürnberg, Germany, E-mail: m.siber@fz-juelich.de}}\par
{\textit{$^{b}$~Department of Chemical and Biological Engineering, Friedrich-Alexander-Universität Erlangen-Nürnberg, Fürther Straße 248, 90429 Nürnberg, Germany}}\par
{\textit{$^{c}$~Department of Material Science and Engineering, Technion Israel Institute of Technology, Haifa 3200003, Israel}}\par
{\textit{$^{d}$~Department of Physics, Friedrich-Alexander-Universität Erlangen-Nürnberg, Fürther Straße 248, 90429 Nürnberg, Germany}}

\fontsize{10pt}{10pt}\selectfont



\addtocounter{figure}{-1}
\begin{figure}[H]
\centering
  \includegraphics[scale=0.25]{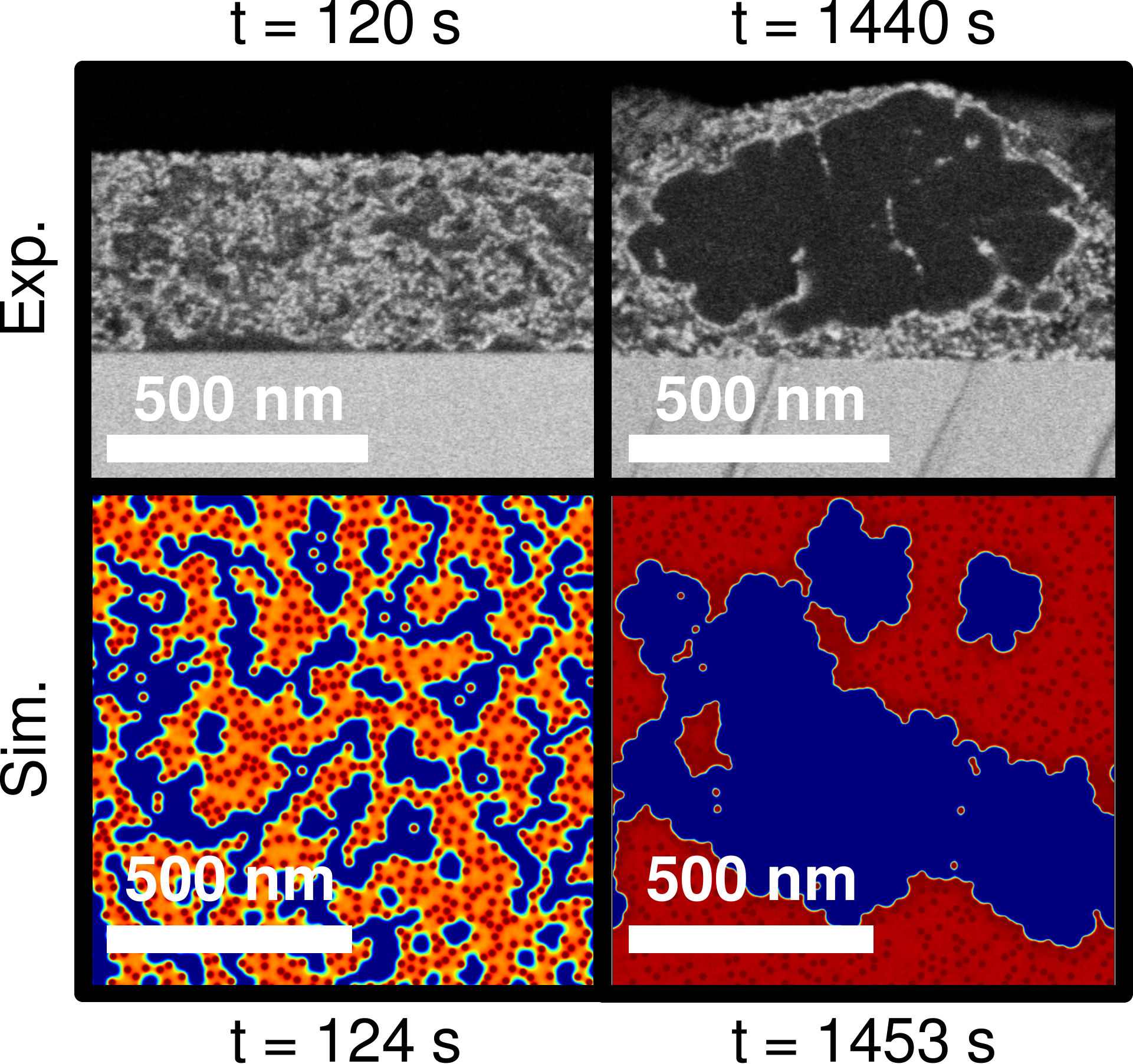}
  \caption{Phase-Field simulations are carried out to elucidate nanomorphology formation mechanisms in organic photoactive layers under thermal annealing. The predicted bulk heterojunction structures are in excellent agreement with Scanning Electron Microscopy measurements. The study provides a detailed comprehension of key mechanistic drivers that need to be understood to devise physically-based strategies for optimized organic electronics fabrication.}
\end{figure}

\section{Abstract}\label{Sec:Abstract}

Tailoring the nanomorphology of organic photoactive layers through a specialized chain of processing steps is an imperative challenge on the path towards reliable and performant organic electronic manufacturing. This hurdle generally proves delicate to be overcome, as organic materials can be subject to many different phase transformation phenomena that are able to interfere with each other and produce a wide variety of morphological configurations with distinct structural, mechanical, and optoelectronic properties. A typical combination of such mechanisms, which the present systems are often prone to, and which is complex to investigate experimentally at the nanoscale, is the phase separation resulting from the interplay between amorphous demixing and crystallization.

In this work, an in-house Phase-Field modeling framework is employed to simulate and, consequently, explain the phenomenological behavior of a photoactive bulk heterojunction during a thermal annealing treatment. The model predictions are validated against available electron microscopy imaging of the nanostructural evolution during the process. It is demonstrated that the simulations can successfully provide a detailed comprehension of crystal nucleation and growth shaped by amorphous spinodal decomposition, so as to yield valuable insights for physically-based morphology control. In addition, this study shows the relevance of extensive thermodynamic and kinetic characterizations (e.g., phase diagram assessments, surface tension measurements, composition-dependent molecular diffusivity evaluations) of organic semiconductor mixtures for the associated field of research.

\section{Introduction}\label{Sec:Introduction}

Owing to their flexible, lightweight, and energy- plus cost-efficient production properties, solution-processed organic semiconductors present attractive perspectives for a broad variety of technology applications~\cite{li_flexible_2018-1,chow_organic_2020,dimov_semiconducting_2022,woo_advances_2023,wang_designing_2024}. In order to realize this potential and achieve a successful introduction to the market, remaining challenges regarding overall performance, long-term stability, industrial upscaling, and non-toxic processing still need to be addressed~\cite{brabec_material_2020,bernardo_progress_2021,mcculloch_sustainability_2023,ding_stability_2024-1,yi_advantages_2024-1, ma_sustainable_2025}.

The core of organic optoelectronic devices, such as organic solar cells (OSCs), organic photo-detectors (OPDs), or organic light-emitting diodes (OLEDs), is the so-called active layer where photons are absorbed, or emitted, and free electric charge carriers are generated, or, conversely, recombined~\cite{wadsworth_bulk_2020}. Depending on the functionality, the employed materials and fabrication steps may differ from one semiconductor type to another. One underlying aspect is, however, shared among all: the nanostructure of solution-cast active layers is determinant to fulfill the above-mentioned requirements for commercialization. Thus, understanding how processing conditions for organic film deposition affect the final layer morphology is of paramount importance in the associated field of research.

This is generally not straightforward, as organic photoactive film formation involves the interplay of complex physical mechanisms at the nanoscale. Typical phase transformations that these types of materials are prone to undergo are driven by concurring amorphous demixing through spinodal decomposition and partial crystallization, which is a nucleation and growth process~\cite{ye_miscibilityfunction_2018,ghasemi_delineation_2019,wang_coupling_2021,peng_materials_2023, zhang_real-time_2025-1}. Multiple experimental techniques (for instance, direct observations with Electron Microscopy or Atomic Force Microscopy, or indirect ones with Ultraviolet-Visible, or X-Ray Spectroscopy) are utilized to assess their effect on morphology. However, deriving a complete and clear picture of the progress of morphology formation during processing is usually impaired due to the various time scales and the relatively small length scales on which the structuring phenomena manifest. This comes in addition to the often similar intrinsic chemical properties of organic molecules, which frequently make them difficult to distinguish from each other. As a consequence, organic active layer optimization is still mainly achieved by trial-and-error~\cite{yi_advantages_2024-1}.

A Phase-Field modelling framework has therefore recently been designed to improve the comprehension of the present process-structure relationships~\cite{ronsin_phase-field_2022}. With this approach, numerical simulations can be carried out to visualize the interplay of distinct physical processes that take place during active layer deposition and post-processing~\cite{ronsin_role_2020,ronsin_phase-field_2020,ronsin_phase-field_2021,konig_two-dimensional_2021,ronsin_formation_2022,siber_crystalline_2023, ameslon_phase_2025-1}. In particular, the model allows to elucidate how crystallization and phase separation phenomena like spinodal decomposition can couple and/or compete~\cite{ronsin_role_2020,siber_crystalline_2023,ameslon_phase_2025-1}.

In a previous study, a parameter space exploration permitted the prediction of distinct morphology formation regimes that arise depending on the balance between thermodynamic and kinetic factors related to both mechanisms~\cite{siber_crystalline_2023}. In this work, the Phase-Field framework is employed to analyze the evolution of the nanostructure in a practical test case. The aim is to evaluate how the developed model can be applied to real systems, which experimental information is necessary upfront to initialize the simulations, how well their predictions agree with experimental observations, and which complementary insights they can procure.

Among the many possible material blends and manufacturing procedures used to produce organic semiconductors, the thermal annealing of a poly[(5,6-difluoro-2,1,3-benzothiadiazol-4,7-diyl)-alt-(3,3’’’-di(2-octyldodecyl)-2,2’;5’,2’’;5’’,2’’’-quaterthiophen-5,5’’’-diyl)]:[6,6]-phenyl-C61-butyric acid methyl ester (PCE11:PCBM) bulk heterojunction is selected as the setup to be investigated. Several reasons motivate this choice: 

\begin{enumerate}
    \item Solar cells relying on this active layer are shown to be subject to severe burn-in performance degradation, which is ascribed to a strong immiscibility between the organic donor and acceptor molecules~\cite{li_abnormal_2017,zhang_comprehensive_2019-1}. Shedding light on the physical mechanisms that drive the resulting phase separation and pinpointing the essential material properties that promote (or hinder) its progress is of significant relevance to improve device stability, not only for this specific bulk heterojunction, but also for a broad class of comparable mixtures employed in organic electronics.
    \item The PCE11:PCBM photoactive layer is a binary blend that fits within the scope of mixtures examined previously with the Phase-Field framework~\cite{ronsin_role_2020,siber_crystalline_2023,ameslon_phase_2025-1}. Moreover, thermal annealing is representative as well of the isothermal situations discussed in the earlier theoretical works~\cite{ronsin_role_2020,siber_crystalline_2023,ameslon_phase_2025-1}. Since it is a commonly performed processing step in organic semiconductor fabrication, gaining knowledge about its effects on active layer morphology is decisive for overall performance optimization.
    \item For this particular material system, reports dedicated to the screening of fundamental blend characteristics can be found in the literature~\cite{levitsky_toward_2020,perea_introducing_2017-1}. These provide an appropriate starting point for deriving the most essential thermodynamic parameters that the Phase-Field model requires. Unfortunately, this crucial information is often incomplete for other material combinations, and sometimes even entirely missing.
    \item A detailed experimental assessment of the morphological evolution of a PCE11:PCBM bulk heterojunction under thermal annealing has also already been completed by Levitsky et al.~\cite{levitsky_bridging_2021}. Especially, the progress of the occurring phase transformations has been imaged using Scanning Electron Microscopy (SEM) on samples that were vapor-phase-infiltrated (VPI) with a selective-staining agent to improve the contrast between different domains. This is of particular interest for the current objectives, as it enables direct comparisons with the phase fields computed in the simulations. 
    \item The pathway for the morphology formation suggested by this latter SEM monitoring~\cite{levitsky_bridging_2021} involves complex phase change phenomena that are indeed driven by the interplay of spontaneous amorphous demixing and crystallization. A comparative study with the Phase-Field model constitutes an effective trial to validate the hypothesized phenomenology and complement its mechanistic description.
\end{enumerate}

Following this introduction (Sec.~\ref{Sec:Introduction}), Sec.~\ref{Sec:ParamSetup} begins with the extraction of thermodynamic input parameters from published experimental data. For a detailed explanation of the implemented computational framework, the reader is referred to prior publications~\cite{ronsin_phase-field_2022, ronsin_formation_2022, siber_crystalline_2023}, which contain extensive material on this topic. Additionally, the model equations relevant to the present investigations are summarized in the supplementary information (SI-A and SI-B). After the parameter setup, first simulations of the thermal annealing process are conducted in Sec.~\ref{Sec:SimpleSims}. The predicted morphology evolution is already found in good agreement with the SEM observations, although a few particular features remain to be explained. Further complexity is thus added to the calculations in Sec.~\ref{Sec:ComplexSims} to capture the missing aspects and complete the analysis. At the end of the manuscript, Sec.~\ref{Sec:Conclusion} summarizes the conclusions and outlooks drawn from this research.

\section{Model Parameter Setup}\label{Sec:ParamSetup}

As mentioned in the introduction, exhaustive thermodynamic characterizations of the PCE11:PCBM system were carried out and reported in previous publications~\cite{levitsky_toward_2020,perea_introducing_2017-1,li_abnormal_2017}. Most importantly, a complete phase diagram could be established for this binary blend from a series of Differential Scanning Calorimetry (DSC) thermograms acquired for numerous mixing ratios~\cite{levitsky_toward_2020}. A substantial benefit from the DSC measurements is that they also yield both the melting temperatures of the components and their latent heats of fusion, which are crucial parameters to model crystallization~\cite{ronsin_phase-field_2022,ronsin_formation_2022,siber_crystalline_2023} (see SI-A and SI-C). 

\begin{figure}[H]
    \centering
    \includegraphics[scale=0.75]{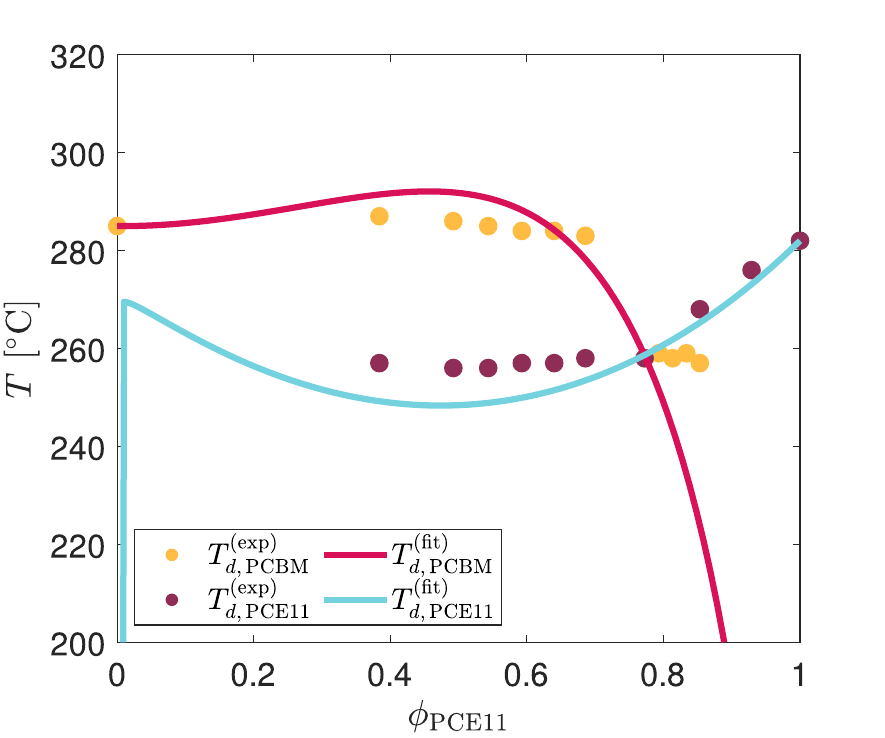}
    \caption{Melting point depression fit (lines) for the binary PCE11:PCBM system according to the Flory-Huggins theory (see Eq.~\ref{eq:MeltingPointDepression}). The experimental data (dots) stems from Differential Scanning Calorimetry (DSC) measurements by Levitsky et al.~\cite{levitsky_toward_2020}. Parameters relevant for the fit are specified in SI-C.}
    \label{fig:MeltingPointDepression}
\end{figure}

Relying on the knowledge of the phase diagram, the melting point depression formula from the Flory-Huggins theory~\cite{flory_principles_1953} can be fitted to the solubility limits of both species (Fig.\ref{fig:MeltingPointDepression}). This permits assessing the value of the interaction parameter that controls the miscibility of the mixture in the amorphous state. The equation is used here in its most general form for both components, rather than with the common simplifying approximations that make the distinction between polymers and small molecules~\cite{nishi_melting_1975}. It follows from the equality of the chemical potentials in the crystalline regions (which are assumed perfectly pure) and the remainder of the mixed amorphous phase, and reads as

\begin{equation}\label{eq:MeltingPointDepression}
            \displaystyle \mu^{(c)} = \mu^{(a)} \Leftrightarrow \frac{\Delta h}{R} \left(\frac{1}{T_{m}} - \frac{1}{T_{d}} \right) =  \ln{\left(\phi^{(a)}\right)}+ \left(1-\phi^{(a)}\right)\left(1- N^* \right) + \chi^{(aa)} N  \left(1-\phi^{(a)}\right)^2  ~.
\end{equation}

Here, $\mu^{(c)}$ and $\mu^{(a)}$ stand for the chemical potentials of the considered component (i.e., either PCE11 or PCBM) in the crystalline and amorphous phase, respectively. $T_m$ denotes the equilibrium melting temperature of the pure material and $T_d$ its measured depression in the blend. $\Delta h$ is the species' molar enthalpy of fusion, and $R$ is the ideal gas constant. $\phi^{(a)}$ represents the volume fraction of the component in the amorphous phase (i.e., the liquidus composition). $N$ is the species size on the Flory-Huggins lattice, and $N^*$ is the ratio of $N$ to the size of the other blend constituent. Finally, $\chi^{(aa)}$ is the aforementioned Flory-Huggins interaction parameter.

Knowing the melting temperatures for different blend ratios as well as the heats of fusion, $\chi^{(aa)}$ is the only degree of freedom left that can be calibrated in order to match both calculated liquidi of PCBM and PCE11 to the experimentally-assessed phase diagram, as depicted in Fig.~\ref{fig:MeltingPointDepression}. The resulting value obtained for the interaction parameter is in line with previous work by Perea et al.~\cite{perea_introducing_2017-1}. Additional comments on this are provided in the SI (SI-C), along with further technical considerations regarding the effective values of the melting temperatures and heats of fusion used for the fit.

Apart from material characteristics such as the molar masses, the densities, or the molar volumes of the blended species, the parameters considered in Eq.~\ref{eq:MeltingPointDepression} are the most critical to properly represent the PCE11:PCBM bulk heterojunction with the current Phase-Field model. Other thermodynamic inputs needed by the model are discussed in the SI (SI-C). A particular remark can be made concerning the interfacial energy of PCBM, which is shown to be relatively strong as compared to other organic small molecules~\cite{bjorstrom_multilayer_2005,barr_nanomorphology_2021,laval_toward_2023}. The values of the corresponding PCBM crystal surface tension parameters are set accordingly high, and have a major influence on the formation of the film morphology (i.e., nucleation density, crystal growth rate, crystal cluster arrangements, as further detailed in Sec.~\ref{Sec:SimpleSims} and Sec.~\ref{Sec:ComplexSims}).

Additionally, the computational framework requires several kinetic parameters to accurately reproduce the time evolution of the system. These regulate the timescales of both diffusive mass transport and crystal nucleation and growth within the blend (see SI-C for more details). No former experimental studies could be found in the literature in order to infer them directly for the PCE11:PCBM mixture. Thus, they are initially estimated from anterior Phase-Field investigations on related material combinations~\cite{ronsin_formation_2022}. Anticipating the content of the next sections (Sec.~\ref{Sec:SimpleSims} and Sec.~\ref{Sec:ComplexSims}), it can be mentioned that the screening of the different stages of the structural evolution under thermal annealing by Levitsky et al.~\cite{levitsky_bridging_2021} (Fig.~\ref{fig:SEMImages}) provides data relative to the timescales of the occurring physical processes, thereby permitting to adjust the kinetic parameters, so that information about their exact value beforehand is not a drastic requirement. The total set of parameters utilized in the upcoming Phase-Field simulations is comprehensively documented in the SI (SI-C).

\section{Simulation of Demixing-Assisted PCBM Crystallization}\label{Sec:SimpleSims}

\begin{figure}[H]
    \centering
    \includegraphics[scale=0.25]{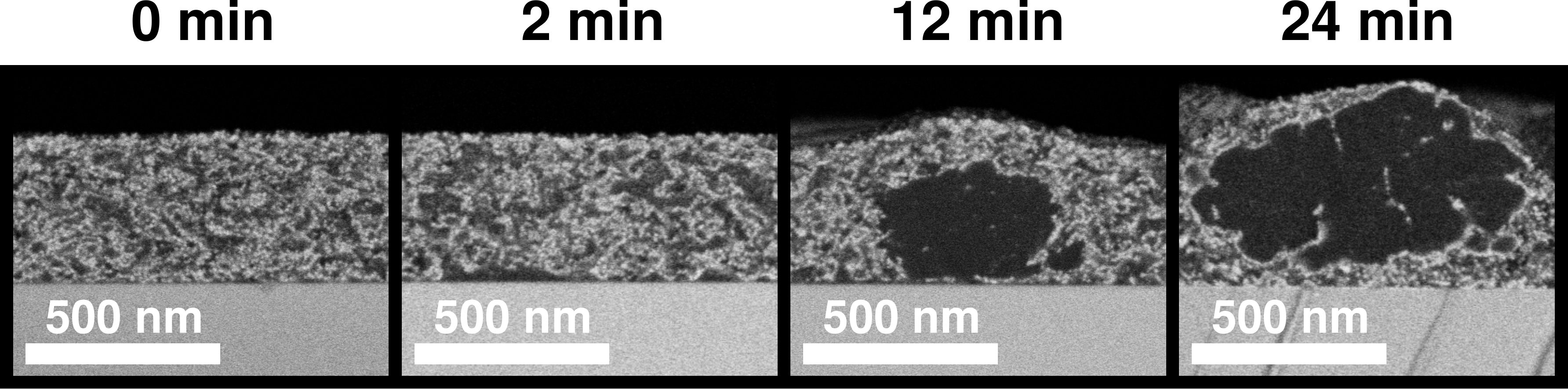}
    \caption{Evolution of PCE11:PCBM morphology upon thermal annealing at 130 $^\circ$C, as imaged with Scanning Electron Microscopy (SEM) by Levitsky et al.~\cite{levitsky_bridging_2021}. The staining agent selectively infiltrates PCE11 domains, which appear bright. PCBM-rich regions range from dark gray (amorphous) to black (crystals). The experiment shows that the mixture is subject to an initial amorphous demixing (see snapshots at 0 and 2 minutes of annealing time), consecutively followed by PCBM crystal nucleation, growth, and clustering (e.g., after 12 and 24 minutes of annealing time). Adapted from Ref.~\cite{levitsky_bridging_2021} with permission from the Royal Society of Chemistry.}
    \label{fig:SEMImages}
\end{figure}

The primary objective of this work is to conduct Phase-Field simulations that accurately predict nanostructural changes within the studied material system under thermal annealing. The morphological evolution of the PCE11:PCBM bulk heterojunction imaged by Levitsky et al.~\cite{levitsky_bridging_2021} at different stages of the treatment is reproduced in Fig.~\ref{fig:SEMImages}. The main steps evidenced on the pathway towards the final active layer structure can be summarized as follows:

\begin{enumerate}
    \item In the as-cast film (0 min), the PCE11 polymer and PCBM small molecule are already incompatible, even in the amorphous state. Thus, an amorphous phase separation process (identified as spinodal decomposition due to the presence of spontaneously appearing elongated domains) immediately starts.
    \item Subsequently, the early stages of the thermal annealing (here before 2 minutes~\cite{levitsky_bridging_2021}, approximately) are characterized by the transformation of both donor- and acceptor-rich amorphous phases. As the phase dissociation proceeds, the associated domains coarsen slightly (up to a length scale of roughly 100 nm~\cite{levitsky_bridging_2021}) and purify in terms of their respective predominant component.
    \item At intermediate annealing stages, PCBM crystals start to nucleate inside PCBM-rich domains. From 6 minutes onwards, a peak, which is associated with PCBM crystallization, becomes observable in the Grazing-Incidence Wide Angle X-Ray Scattering (GIWAXS) diffraction pattern~\cite{levitsky_bridging_2021}. In the SEM images, the presence of crystals is detectable in PCBM domains, which exhibit deeper contrast with the bright PCE11 regions, indicating that further density changes (attributed to the crystallization of PCBM) occur in addition to the purification undergone during the amorphous phase separation.
    \item The PCBM crystallization phase transition is strongly growth-dominated, that is, a low nucleation density is observed, so that materializing crystal grains also have time to grow and available space to fill before they impinge with each other. As a result, crystals become significantly larger than the amorphous domains they originated from (up to 500 nm in diameter). The growth process is globally isotropic, even though it can be seen that the shape of the interfaces between crystals and their amorphous surroundings tend to present irregularities (see Fig.~\ref{fig:SEMImages} at 12 min).
    \item On larger timescales (24 minutes), PCBM crystals group in micrometer-sized clusters. Interfaces between grain boundaries are visible with a bright contrast, which implies a higher PCE11 content at these locations. Moreover, the top surface of the film tends to present pronounced bulges at cluster positions, likely due to the strong surface tension of PCBM crystals. This latter bulging effect is not sought to be reproduced here in the simulations, as its investigation requires the inclusion (among other additional features) of the ambient air in the considered system, which is outside the scope of this study.
    \item At 40 minutes of thermal annealing, the Photoluminescence (PL) signal, which tracks the presence of the amorphous phase separation, is saturated~\cite{levitsky_bridging_2021}, suggesting that the crystallization of PCBM consumed all of the PCBM-rich amorphous regions, so that solely a single amorphous phase remains with a predominant content of PCE11 material. Around the same time (30 minutes), the GIWAXS PCBM peak stops evolving as well, indicating that the crystal growth has ceased.
    
\end{enumerate}

Based on these experimental observations, Levitsky et al. hypothesized a schematic phase diagram explaining the effective transformations undergone by the PCE11:PCBM active layer under thermal annealing~\cite{levitsky_bridging_2021}. As compared to the melting point data discussed in the previous section (Fig.~\ref{fig:MeltingPointDepression} in Sec.~\ref{Sec:ParamSetup}), the suggested diagram omits the region pertaining to PCE11 crystallization. This is supported by the evidence that no notable changes are observed during the whole thermal annealing process in the time-resolved PL and GIWAXS signals attributed to PCE11 crystals~\cite{levitsky_bridging_2021}. The main feature of the diagram is an amorphous miscibility gap, which justifies that the phase separation in the as-cast film occurs via spinodal decomposition.

\begin{figure}[H]
    \centering
    \includegraphics[scale=0.8]{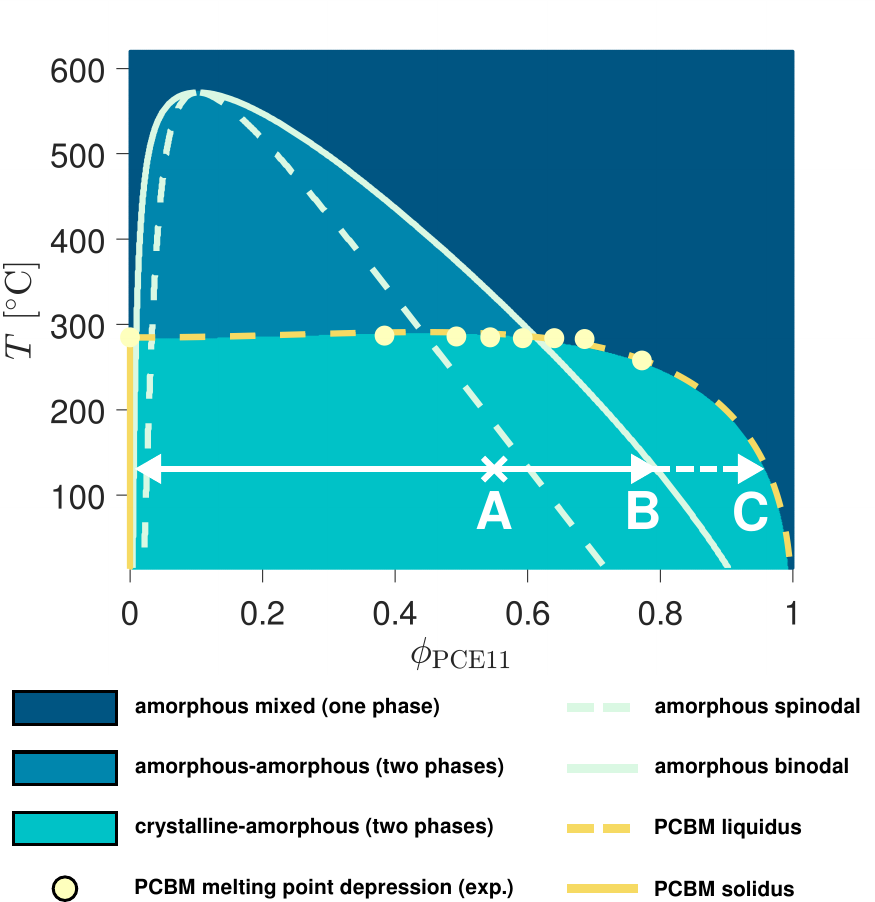}
    \caption{Phase diagram calculated for the PCE11:PCBM system using the thermodynamic parameters calibrated with DSC experiments (see Sec.~\ref{Sec:ParamSetup}). PCE11 crystallization is omitted according to the interpretation of the thermal annealing process by Levitsky et al.~\cite{levitsky_bridging_2021}. The evolution of the mixture composition at 130~$^\circ$C is specified by the white arrows. The as-cast blend first presents a single unstable amorphous phase with 55~v\% of PCE11 (point A). Spontaneous amorphous demixing via spinodal decomposition generates two phases, respectively enriched in PCE11 and PCBM. The composition of the phases is driven towards the binodal (point B for the PCE11-rich one). Subsequently, PCBM crystallization induces further phase content purification until the crystalline-amorphous equilibrium is attained (point C denotes the corresponding liquidus composition).}
    \label{fig:SimpleTADiagram}
\end{figure}

Before performing numerical simulations, a quantitative phase diagram can be calculated for the PCE11:PCBM blend from the free energy of the Phase-Field model. Among the available methods, the convex hull approach~\cite{hildebrandt_predicting_1994-1,voskov_ternapi_2015,mao_phase_2019,gottl_convex_2023} is selected here to do this. Relying on the thermodynamic parameters calibrated in Sec.~\ref{Sec:ParamSetup} with the melting point depression fits (and neglecting the contribution related to PCE11 crystallization), the obtained diagram (Fig.~\ref{fig:SimpleTADiagram}) notably agrees with the interpretation of Levitsky et al.~\cite{levitsky_bridging_2021}. Most importantly, the fitted value for the Flory-Huggins interaction parameter indeed predicts amorphous incompatibility over a large range of mixture compositions and temperatures.

Following this, Phase-Field simulations of active layer formation are carried out for the mixture described by this simplified phase diagram (Fig.~\ref{fig:SimpleTADiagram}) at the blend ratio and the temperature previously investigated in the experiment of Levitsky et al.~\cite{levitsky_bridging_2021} (0.45 w\%, that is approximately 0.55 v\% of PCE11 at T = 130~$^\circ$C). The main idea is to compare their outcome against the morphology formation pathway outlined above and the SEM images displayed in Fig.~\ref{fig:SEMImages}. 

\begin{figure}[H]
    \centering
    \includegraphics[scale=0.24]{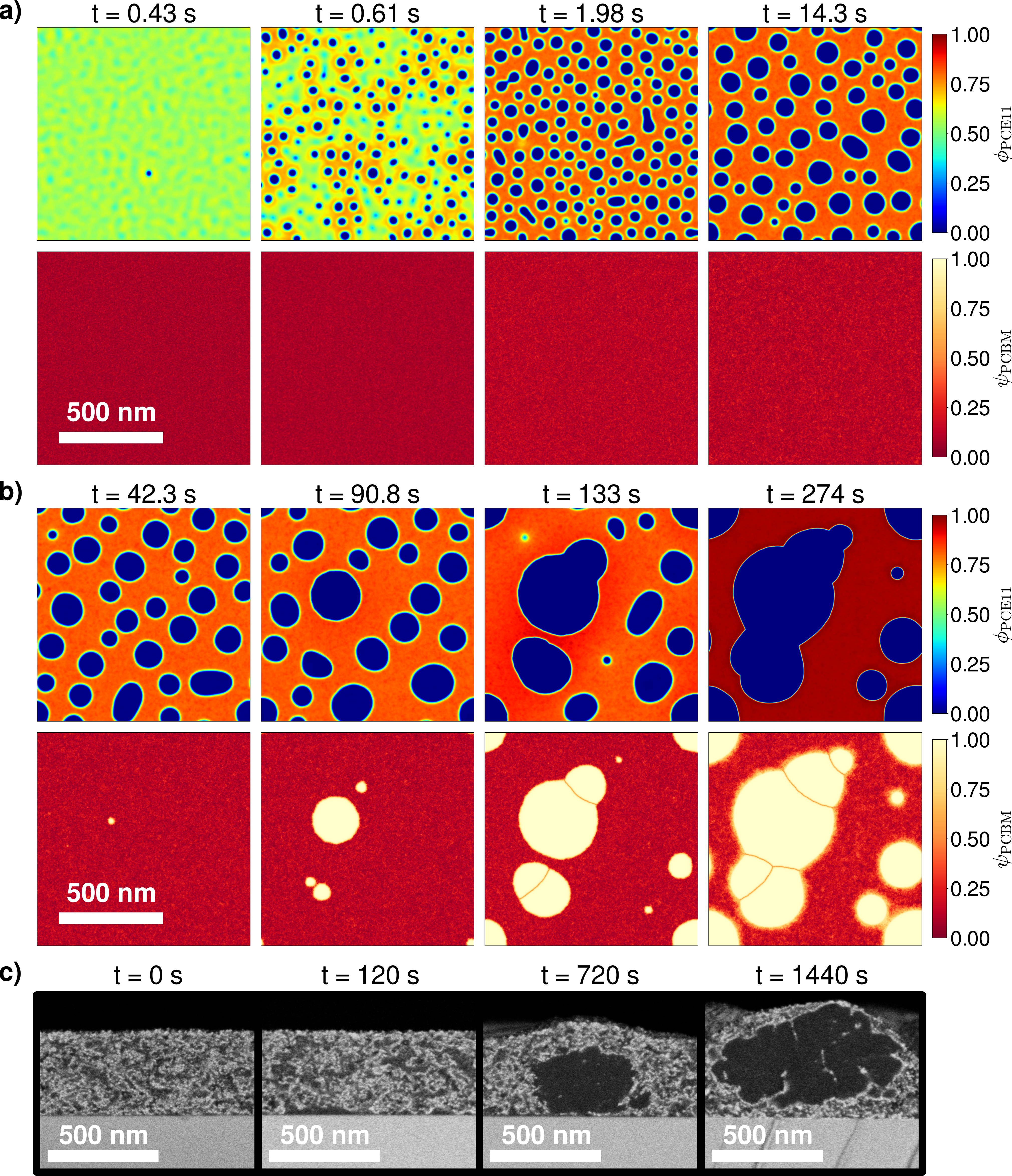}
    \caption{Simulated morphology evolution of the PCE11:PCBM mixture under thermal annealing at 130$^\circ$C. a) Early annealing stages with initial amorphous demixing. b) Intermediate to late annealing stages on which growth-dominated PCBM crystallization takes place. In both a) and b), the first row presents the volume fraction field of the PCE11 polymer ($\phi_{\mathrm{PCE11}}$). The second row then displays the PCBM order parameter field ($\psi_{\mathrm{PCBM}}$) which monitors the presence of PCBM crystals ($\psi_{\mathrm{PCBM}}=1$) in the system. The observed phase transformations arise according to the phase diagram represented in Fig.~\ref{fig:SimpleTADiagram}. All relevant simulation parameters are specified in the SI (SI-C). For comparison with the experiments, the SEM acquisitions of Levitsky et al.~\cite{levitsky_bridging_2021} are also reproduced in c) (Adapted from Ref.~\cite{levitsky_bridging_2021} with permission from the Royal Society of Chemistry).}
    \label{fig:SimpleTA}
\end{figure}

To confirm the nature of the phase separation process visible in the snapshots of the as-cast film, the system is initialized as a fully mixed amorphous phase, in which both components are present in the proportions of the overall blend ratio. It can be noted that the simulated domain is a two-dimensional 1024$\times$1024 nm square box with periodic boundary conditions along each axis, thereby representing a horizontal cross-section of the bulk of the active layer. This is nonetheless also representative of the vertical cross-sections visualized with SEM, as Levitsky et al.~\cite{levitsky_toward_2020} demonstrated with 3D tomography that the structures arising from the phase dissociation between PCE11 and PCBM are isotropic. As showcased in the SI, a large variety of morphologies can be obtained depending on kinetic and surface tension parameters, even when the thermodynamic properties that determine the phase diagram remain identical (SI-D, SI-E, SI-F). Fig.~\ref{fig:SimpleTA} presents the simulation case that matches best with the features observed experimentally.

Directly at the start of the simulation, the blend spontaneously demixes (t = 0.43~s in Fig.~\ref{fig:SimpleTA}). The spinodal decomposition is identified by the emergence of a wave pattern shape that is characteristic of this phenomenon. The length scale of the wave pattern is dictated by the molecular interactions (described by the Flory-Huggins interaction parameter), the surface tension, and the blend ratio~\cite{konig_two-dimensional_2021}. A fast purification of the dissociated phases proceeds, which agrees with the timescale of the SEM acquisitions~\cite{levitsky_bridging_2021}. Due to this ongoing uphill diffusion of both materials into their respective phases, and under the influence of surface tension, the wave pattern breaks down into a droplet-in-background-matrix structure (t = 0.61~s). The predominant constituents in the droplets and in the background are primarily determined by the blend ratio (see SI-D), but can also be affected by composition-dependent diffusion properties. In the present case, the droplets and the matrix are respectively PCBM-rich and PCE11-rich. 

Once the concentrations foreseen by the corresponding binodal curves (Fig.~\ref{fig:SimpleTADiagram}) are reached in both phases (e.g., t = 1.98~s), the progress of the amorphous phase separation becomes mainly driven by Ostwald ripening: In order to minimize the amount of energetically detrimental interfaces between domains of different compositions, the system favors the growth of larger droplets at the expense of smaller ones, thus effectively reducing the droplet surface-to-volume ratio (t = 14.3~s). Ostwald ripening is a significantly slower process than the prior morphological transformations. During the timescales of this stage, thermal fluctuations are likely to trigger PCBM crystal nucleation within the amorphous PCBM droplets (first event at t = 42.3~s). Since the thermodynamic driving force for PCBM crystallization is enhanced in regions with high PCBM content, the crystallization phase transition is accordingly accelerated in the droplets. The crystallization process is promoted by the initial amorphous-amorphous phase separation and can therefore be categorized as demixing-assisted~\cite{siber_crystalline_2023}. 

A substantial morphology-shaping property of the PCBM crystals is their strong surface tension. This relates to a high energy barrier for crystallization and a rather wide critical size to be overcome by stochastically materializing ordered PCBM regions before they effectively form a stable crystal seed that does not collapse back to the amorphous state. Thus, the overall probability of forming a new nucleus is still relatively low. Because of this, there is more material and space available for already formed germs to expand and grow large (t = 90.8~s), resulting in a growth-dominated crystallization regime. Amorphous PCBM molecules incorporated into a crystal do not all originate from the PCBM domain in which the nucleus initially appeared, as they can also diffuse from neighbouring droplets through the PCE11-rich phase to attach to the growth front. As a result, the PCBM crystals grow larger in size than the droplets they are issued from (t = 133~s). This corresponds to the crystallization behavior seen in the SEM images. Moreover, the uptake of PCBM material from surrounding amorphous droplets causes their depletion until they are completely consumed (t = 274~s), which is in line with the analysis of the PL data performed by Levitsky et al.~\cite{levitsky_bridging_2021}.

Outgoing from of the reference experimental study~\cite{levitsky_bridging_2021}, it is still open whether the PCBM crystal growth is diffusion-controlled as in other related organic photovoltaic systems~\cite{berriman_molecular_2015}, or not. The simulations suggest that the system is not strongly diffusion-limited here, since, in the latter case, growing crystals tend to be separated from each other by areas depleted in PCBM~\cite{siber_crystalline_2023} (see SI-F), which prevent them from impingement. In non-diffusion-limited crystallization scenarios, such as Fig.~\ref{fig:SimpleTA}, PCBM crystals ultimately impinge when they become large, leading to the formation of crystal clusters at late annealing stages (t = 274~s). Clusters are more favorable than a dispersed crystal arrangement for the global free energy balance of the system due to the strong surface tension of the PCBM crystals in an amorphous environment.

All in all, the morphology formation pathway realized in this simulation captures most of the phenomenology observed in the SEM experiments. Most importantly, it validates the mechanistic hypotheses of Levitsky et al.~\cite{levitsky_bridging_2021}: The nanostructural changes are indeed driven by the combination of a spontaneous amorphous phase separation through spinodal decomposition and a subsequent demixing-assisted PCBM crystallization transition that follows a nucleation and growth process. Thus, it can be confirmed that both mechanisms (i.e., spinodal decomposition and nucleation and growth) are not to be regarded as mutually exclusive. To clarify this, it is necessary to distinguish between the nature of the phase change phenomenon (e.g., demixing or crystallization) and the phase change mechanism (i.e., spinodal decomposition or nucleation and growth) through which it occurs. In this way, amorphous demixing can occur either through spinodal decomposition as in Fig.~\ref{fig:SimpleTA} if the mixture is unstable, or through nucleation and growth, in case it is metastable (i.e., when the blend ratio lies within the region between the spinodal and binodal lines of the phase diagram - see Fig.~\ref{fig:SimpleTADiagram}). In comparison, crystallization proceeds in general by nucleation and growth, even though spinodal crystallization may be envisaged as well (at least theoretically, for thermodynamic systems described by a free energy formulation with a vanishing energy barrier below a certain temperature threshold~\cite{granasy_phase-field_2019-1}). The interplay of amorphous phase separation and crystallization can therefore imply:

\begin{enumerate}
    \item Amorphous phase separation via spinodal decomposition, followed by demixing-assisted crystallization by nucleation and growth, like in the presently investigated PCE11:PCBM bulk heterojunction.
    \item Amorphous phase separation via nucleation and growth, followed by demixing-assisted crystallization by nucleation and growth, which is in principle also possible for the PCE11:PCBM system at a different blend ratio.
    \item Crystallization by nucleation and growth, followed by amorphous phase separation via spinodal decomposition. In this situation, the initial crystallization process is diffusion-limited, as it occurs ahead of the spinodal decomposition, whose kinetics are determined by the diffusion properties of the blend. Additionally, the amorphous phase separation unfolds around the crystals in specific geometric patterns, as detailed in previous Phase-Field studies~\cite{ronsin_role_2020, siber_crystalline_2023}.
    \item Crystallization by nucleation and growth, followed by amorphous phase separation via nucleation and growth. In this case, the crystallization can act as a trigger for the formation of amorphous droplets (see also earlier theoretical work by Siber et al.~\cite{siber_crystalline_2023}).
\end{enumerate}

While the current simulation provides a detailed explanation of the successive physical processes undergone by the PCE11:PCBM bulk heterojunction during the thermal annealing, two specific aspects of the observed morphologies remain to be clarified: First, the global shape of the amorphous PCBM regions in which crystals materialize differs between the Phase-Field prediction (round droplets in Fig.\ref{fig:SimpleTA}) and the SEM observations (irregular, elongated, and sometimes branching structures with dented interfaces in Fig.\ref{fig:SEMImages}). Second, it is not possible, with the present modeling assumptions, to match quantitatively the time and length scales of the computations with those of the experiment. Indeed, the diffusion properties of the blend are calibrated, so that the development of the initial demixing agrees with the SEM investigation. However, this also settles the timescales on which the subsequent Ostwald ripening takes place. With the utilized material diffusivities, the size of the coarsening amorphous PCBM domains then exceeds significantly the experimental estimation at the time where PCBM nucleation is expected to happen (below 100 nm in width around 6 min). Thus, the model parameter controlling the onset (along with the global kinetics) of PCBM crystallization is adjusted in the present simulation (Fig.~\ref{fig:SimpleTA}) to obtain nucleation and growth when the PCBM droplets have the adequate size. This nevertheless comes at the cost of an inaccurate timing for the general crystallization process (first nucleus before 1 min, final clusters already around 4 min). It is the purpose of the next section to address both issues and refine the Phase-Field analysis of this case study with more sophisticated simulations.

\section{Morphology-Defining Effect of PCE11 Crystallites}\label{Sec:ComplexSims}

As explained at the beginning of the previous section, PCE11 crystallization is not expected to develop during the thermal annealing since the corresponding parts of the PL and GIWAXS spectra do not reveal any notable change with the treatment~\cite{levitsky_bridging_2021}. Nevertheless, both signals still show that a certain amount of PCE11 crystals is actually contained in the as-cast film, even though it does not evolve afterwards. The mere presence of polymer crystallites within the system is, however, sufficient to substantially impact the progress of morphology formation.

To demonstrate this, further Phase-Field simulations are conducted assuming the presence of PCE11 crystallites in the as-cast photoactive layer (Fig.~\ref{fig:ComplexTA}). For simplicity (and because the exact geometrical features of the PCE11 crystals are not known from the experiments), all PCE11 seeds are set up identically in size and shape, although it is likely that both follow some sort of distribution. This simplification, however, proves satisfactory to model the morphological properties investigated here. The size of the crystallites is fixed according to the coherence length measured by GIWAXS~\cite{levitsky_bridging_2021} (below 20~nm). The total amount of PCE11 crystallites found to yield the best agreement between simulated and SEM morphologies corresponds to an overall extent of about 33\% PCE11 crystallinity. This is expected since PCE11 is designed to exhibit a relatively high crystallinity with respect to other polymers used in organic photovoltaics~\cite{zhang_comprehensive_2019-1,zhang_pce11-based_2020}. The value also matches with the results of Perea et al.~\cite{perea_introducing_2017-1} who calculated a crystalline PCE11 mass fraction of 32\% in pristine PCE11 films based on DSC scan analyses. 

\begin{figure}[H]
    \centering
    \includegraphics[scale=0.8]{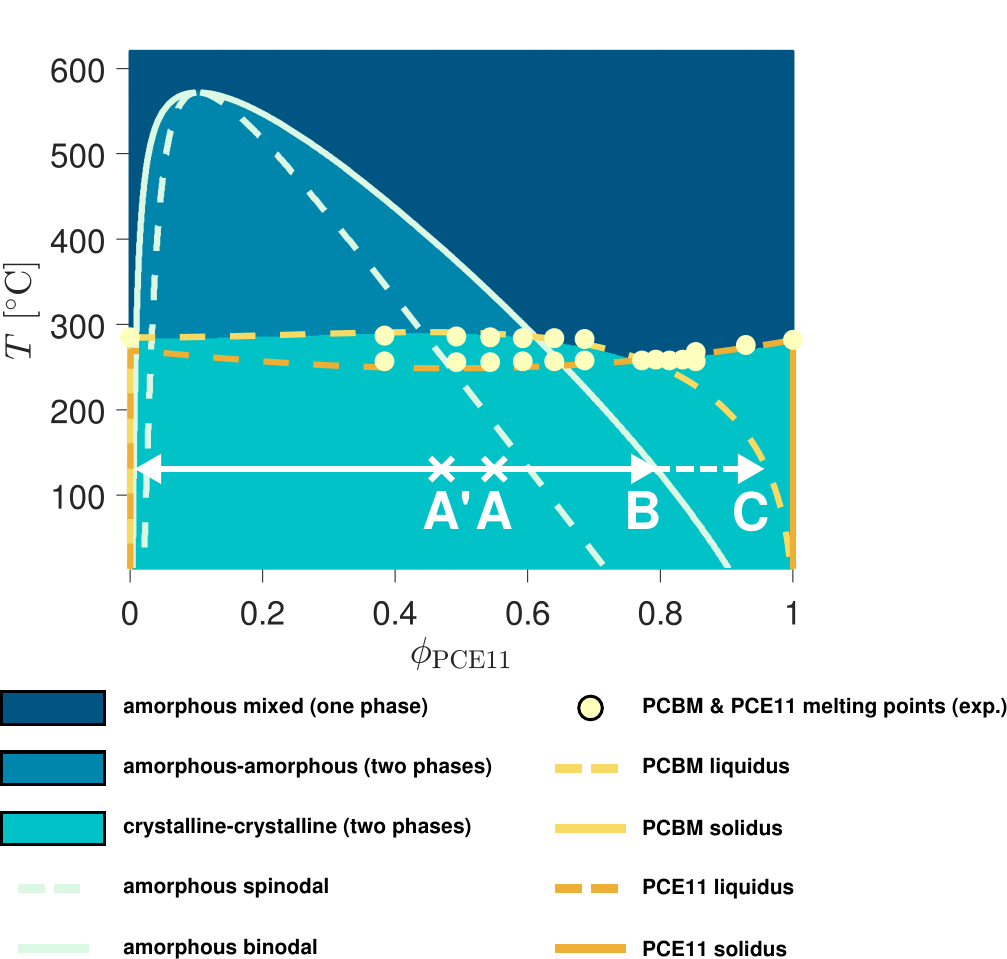}
    \caption{Phase diagram calculated for the PCE11:PCBM system using the thermodynamic parameters calibrated with DSC experiments (see Sec.~\ref{Sec:ParamSetup}). The evolution of the mixture composition at 130~$^\circ$C is specified by the white arrows. The as-cast blend with an overall blend ratio of 55 v\% of PCE11 (point A) already contains pure PCE11 crystallites along with an unstable amorphous phase with 45 v\% of PCE11 (the latter is represented by point A'). In this phase, spontaneous amorphous demixing via spinodal decomposition generates two phases, respectively enriched in PCE11 and PCBM. The composition of these phases is driven towards the binodal (point B for the PCE11-rich one). Subsequently, PCBM crystallization induces further phase content purification until a crystalline-amorphous pseudo-equilibrium is attained (point C denotes the corresponding liquidus composition). Note that this final state with one amorphous (PCE11-rich) and two separate crystalline phases (pure PCBM and PCE11, respectively) is not the global equilibrium predicted for the system.
    The expected equilibrium is only composed of both pure PCBM and PCE11 crystal phases. However, reaching this is impeded by the semicrystalline nature of the PCE11 polymer, which prevents it from exceeding the amount of crystallite content in place at the beginning of the annealing.}
    \label{fig:ComplexTADiagram}
\end{figure}

\begin{figure}[H]
    \centering
    \includegraphics[scale=0.24]{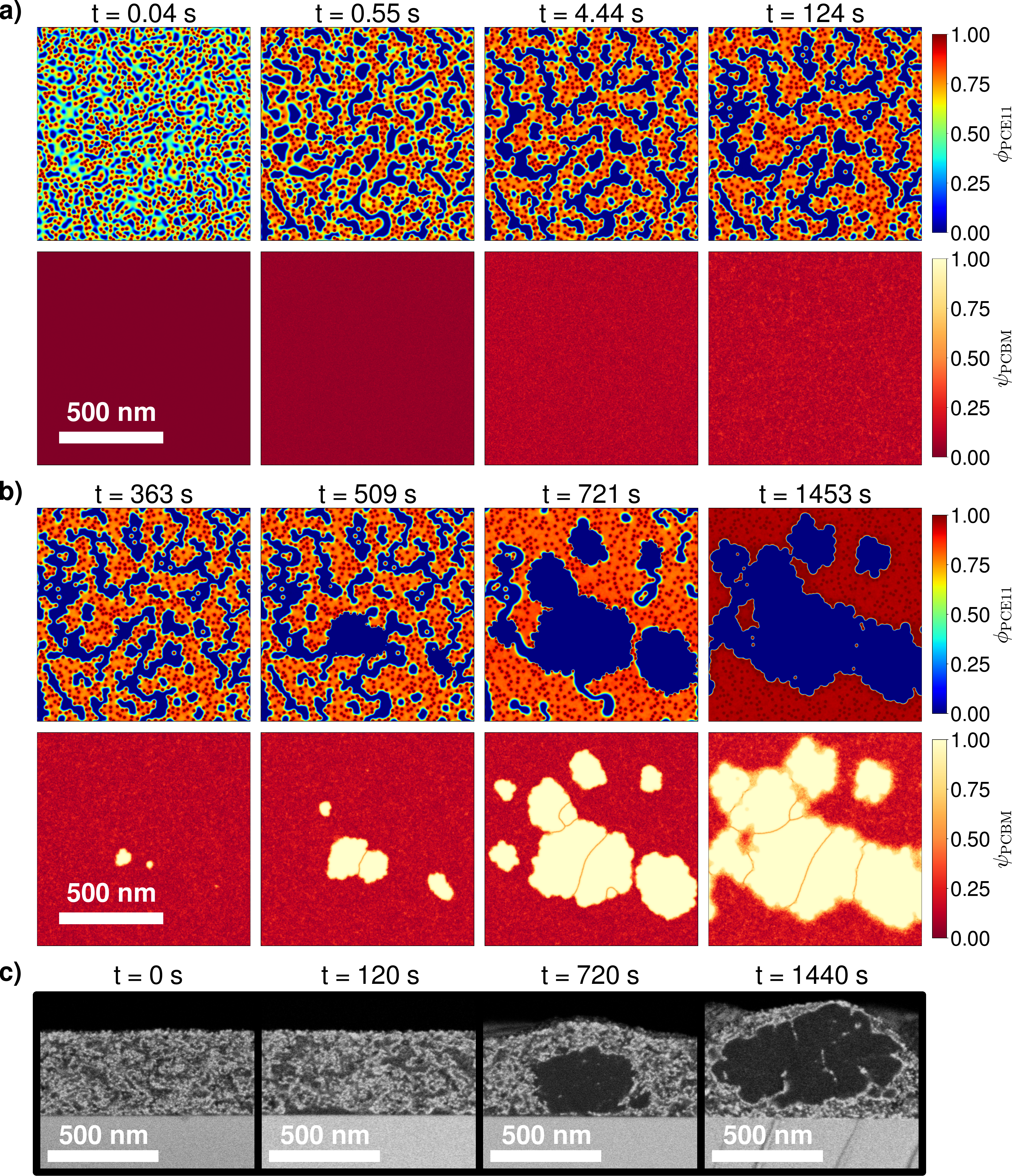}
    \caption{Simulated morphology evolution of the PCE11:PCBM mixture under thermal annealing at 130~$^\circ$C. a) Early annealing stages with initial amorphous demixing shaped by PCE11 crystallites. b) Intermediate to late annealing stages on which growth-dominated PCBM crystallization takes place. In both a) and b), the first row presents the volume fraction field of the PCE11 polymer ($\phi_{\mathrm{PCE11}}$). The second row then displays the PCBM order parameter field ($\psi_{\mathrm{PCBM}}$), which monitors the presence of PCBM crystals ($\psi_{\mathrm{PCBM}}=1$) in the system. The observed phase transformations arise according to the phase diagram represented in Fig.~\ref{fig:ComplexTADiagram}. All relevant simulation parameters are specified in the SI (SI-C). For comparison with the experiments, the SEM acquisitions of Levitsky et al.~\cite{levitsky_bridging_2021} are also reproduced in c) (Adapted from Ref.~\cite{levitsky_bridging_2021} with permission from the Royal Society of Chemistry).}
    \label{fig:ComplexTA}
\end{figure}

The kinetic model parameters determining the PCE11 crystal growth rate are set relatively low (see SI-C), so as to effectively prevent a significant crystalline PCE11 volume increase, as expected from the PL and GIWAXS data. The phase diagram, taking into account both PCBM and PCE11 crystalline phases (Fig.~\ref{fig:ComplexTADiagram}), now incorporates all the thermodynamic information assessed using DSC (Fig.~\ref{fig:MeltingPointDepression}). Since the total blend ratio of the mixture (which includes the material of the relatively pure PCE11 crystals) is the same as before (i.e., 55 v\% of PCE11), the initial concentration of PCE11 in the mixed amorphous phase is consequently reduced (namely to 45 v\% of PCE11).

In general, the morphology evolution (Fig.~\ref{fig:ComplexTA}) follows the same sequence of physical processes as the one comprehensively detailed in the precedent section: spontaneous amorphous demixing through spinodal decomposition, amorphous domain purification, coalescence and coarsening, growth-dominated PCBM crystallization in PCBM-rich regions, PCBM crystal cluster appearance at late stages. Nonetheless, noteworthy specificities arise in the present case, which permit alleviation of both shortcomings noted in the previous simulation (Fig.~\ref{fig:SimpleTA}).

The first major effect of the PCE11 crystal seeds is an alteration of the development of the initial spinodal demixing. Rather than taking place via the expected wave pattern, it unfolds around the crystallites, which already present a locally elevated content of PCE11 material and serve as catalysts for the phase separation (t = 0.04~s in Fig.~\ref{fig:ComplexTA}). Furthermore, the purification of the domains occurs more progressively without a sharp rupture into droplets and background matrix structure (t = 0.55~s), as compared to Fig.~\ref{fig:SimpleTA}. This matches better with the morphological evolution shown in the SEM snapshots of the early annealing stages (see Fig.~\ref{fig:SEMImages}). 

When the binodal composition is reached in the amorphous PCBM phase (t = 4.44~s), further coalescence and coarsening occur. The emerging structures are more elongated and irregular in shape than the droplets of the simplified thermal annealing case (Fig.~\ref{fig:SimpleTA}), and nearly build a co-continuous morphology. This is due, on the one hand, to the initial composition of the mixed amorphous phase, which has a reduced PCE11 concentration and is therefore closer to the material proportions appropriate for bi-continuous percolation (see SI-D). On the other hand, PCE11 crystal seeds obstruct the displacement of amorphous phase interfaces, thus preventing the PCBM-rich domains from relaxing towards energetically more favorable round configurations. The resulting amorphous domain geometry and distribution conform accurately with the practical observations. 

It can be noted that the PCE11 crystallites also quench the coarsening of the amorphous PCBM regions at relatively early stages (i.e., before t = 124~s). Their maximum attainable size is dictated by the PCE11 crystal density. Since further coarsening of the amorphous phases is impeded, the length- and timescales of the initial demixing stages and the PCBM crystallization are effectively decoupled from then on. Simulations where the kinetic parameter that controls PCBM nucleation and growth is varied are shown in the SI (SI-G), highlighting that, beyond the quench timing, the morphological changes arising from the PCBM crystallization are no longer dependent on the timescales on which the phase transition occurs. The quench of the coarsening of the amorphous phase separation by the PCE11 crystallites thus allows PCBM crystallization to be appropriately shifted in time towards the experimentally measured timescales (Fig.~\ref{fig:SEMImages}), while maintaining the width of the remaining amorphous PCBM domains within the observed range (up to around 100~nm) during the whole process.

In Fig.~\ref{fig:ComplexTA}, PCBM nucleation happens on timescales separate from the initial phase segregation (t = 363~s), which agrees with the experimental measurements. The nuclei still materialize in PCBM-rich regions and grow isotropically. Once a growing crystal reaches a domain boundary and impinges on the PCE11 crystallites (t = 509~s), its interface experiences a deformation, hence leading to the irregular PCBM crystal growth fronts that are also seen in the SEM experiments.

The PCBM crystals are, however, not fully hindered from growing, as PCBM molecules continue to diffuse towards their surface from the surrounding amorphous material. As a consequence, the crystal begins to develop around the PCE11 seed. As a PCE11 crystallite becomes progressively encircled by the PCBM crystal phase, the surface tension at the PCBM-PCE11 crystal-crystal interface becomes significantly stronger. The associated free energy rise counterbalances the gain obtained by the crystallization of the PCE11 material, which is gradually less favorable for the system. Eventually, the PCE11 crystallite becomes thermodynamically unstable, leading to its dissolution and the diffusion of its relatively pure PCE11 content into the amorphous PCE11 phase. This allows PCBM crystals to retain a relatively isotropic shape, even at late growth stages (t = 721 s) where their sizes exceed that of the amorphous phase separation pattern and the initial distances between the PCE11 crystallites. 

It is likely that a PCE11 crystal dissolution releases entanglement constraints on the polymer chains formerly involved in the crystallite, so that a new PCE11 crystal seed may reform further away from the PCBM crystal interface in replacement of the dissolved one. Alternatively, this can also allow other existing PCE11 grains to grow slightly larger, so as to globally conserve the total amount of crystallized PCE11. This is not captured in the present simulations, as it requires a more elaborate treatment of the semicrystalline nature of the polymer, which necessitates further theoretical and numerical developments.

The present simulation confirms that the PCBM crystal growth is not strongly diffusion-limited. In the opposite case (see SI-F), PCE11 crystallites at the PCBM growth front do not have sufficient time to dissolve and diffuse away. Consequently, the PCBM seed rather tends to grow around them, resulting in substantially different crystalline structures with more elongated branches, as compared to the overall nearly round crystals with indented interfaces obtained in Fig.~\ref{fig:ComplexTA} and observed with SEM.

PCBM crystal growth is sustained until all amorphous PCBM domains are completely consumed (t = 1453~s). The amorphous PCE11-rich regions purify as well, so that they evolve from the binodal composition to the liquidus one (Fig.~\ref{fig:ComplexTADiagram}). At late stages, large PCBM crystal grains impinge (t = 721~s) and remain agglomerated in clusters due to the strong surface tension of PCBM crystals in an amorphous environment (t = 1453~s). PCE11 crystallites can become trapped at PCBM grain boundaries if they are located near their impingement line. Moreover, the diffusivity in the amorphous polymer-rich phase decreases drastically when it loses its minor PCBM content, so that PCE11 crystallites then persist longer at the growth front and are more likely to be incorporated in between PCBM crystal grains. This is in agreement with the SEM results of Levitsky et al.~\cite{levitsky_bridging_2021} that exhibit narrow bright contrast zones characteristic of a localized high PCE11 content at grain boundaries within PCBM clusters.

Overall, the present simulation shows that taking into consideration the presence of PCE11 crystallites in the as-cast film permits to phenomenologically explain and quantitatively replicate the morphology evolution observed during the thermal annealing. Complementing the previous description by Levitsky et al.~\cite{levitsky_bridging_2021}, the study delivers critical insights regarding the key mechanistic drivers responsible for the nanostructural changes, namely:

\begin{enumerate}
    \item The PCE11 crystallites have a double impact on the morphology, even though they are not actively affected by the applied processing conditions. First, they act as a catalyst for the initial spinodal demixing and define the ensuing domain geometry. Second, they prevent the coarsening of the domains, thereby freezing the amorphous phase separation pattern in a nearly percolating configuration. 
    \item The generated amorphous PCBM-rich domains give rise to an enhanced driving force for PCBM crystallization. Their spatial distribution determines the locations where nucleation can take place. Since the amorphous spinodal decomposition promotes the crystallization, the process is identified as demixing-assisted.
    \item The high surface tension of PCBM crystals is the main reason why the crystallization is growth-dominated. In addition, it causes PCE11 crystallites to dissolve at PCBM-PCE11 crystal-crystal boundaries, which leads to the disruption of the quenched bi-continuous phase separation pattern by the isotropically growing PCBM crystals. Finally, surface tension also favors the crystals aggregating in clusters, as opposed to a dispersed arrangement.
    \item The kinetics of PCBM crystallization are not diffusion-limited here, effectively providing time for the PCE11 crystallite dissolution (which allows the globally isotropic PCBM crystal growth), and enabling the eventual PCBM crystal impingement at late annealing stages (which precedes the cluster formation).
\end{enumerate}

\section{Conclusion}\label{Sec:Conclusion}

In conclusion, the thermal annealing of the organic PCE11:PCBM bulk heterojunction was successfully reproduced with the Phase-Field method. Simulations accounting for all observed notable morphological features were achieved, granting valuable insights into the phenomenology induced by the thermal treatment. The comparison between the predicted morphology evolution and the experimental nanostructure monitoring with VPI-enhanced SEM confirmed that the system is subject to a demixing-assisted crystallization scenario: An initial amorphous phase separation occurring via spinodal decomposition generates PCBM-rich domains where crystal nucleation and growth are facilitated. 

Beyond this, the detailed mechanistic comprehension provided by the Phase-Field analysis allowed to evidence the critical effect of the presence of PCE11 crystallites in the as-cast film. The polymer crystallites were indeed shown to dictate the geometry of the initial amorphous phase segregation and to subsequently stabilize it by quenching the consecutive domain coarsening. On separate annealing timescales, non-diffusion-limited, growth-dominated PCBM crystallization driven by high crystal surface tension was revealed to be responsible for the eventual alteration of the amorphous phase separation pattern, leading to the final morphology consisting of large PCBM crystal clusters surrounded by pure PCE11 material in both amorphous and crystalline form.

Many characteristics of the unraveled mutual influence of crystallization and amorphous phase separation on blend morphology are expected to be valid for organic material combinations beyond the PCE11:PCBM system. This study thus demonstrates the promising potential of the Phase-Field approach to unveil and rationalize determining structuring aspects that must be considered when aiming to adjust performance-relevant morphological properties (such as domain sizes, shapes, and spatial distributions) over a complex sequence of processing steps. With respect to the upscaling and commercialization challenges related to organic electronics, future applications for this type of simulations include the investigation of differences between thin and thick film deposition, the screening of process parameter variations, the comparison of donor-acceptor mixtures cast from halogenated and non-halogenated solvents, the effect of post-treatment strategies, and the nanostructural changes arising during active layer aging.

Finally, while the outcome of this work generally calls for a more extensive elucidation of the fundamental thermodynamic and kinetic properties of the materials used to fabricate organic active layers, the assessment of the phase diagrams proved to yield indispensable and powerful information to accurately apprehend the behavior of the investigated mixtures. It is therefore the aim of the present manuscript to encourage further analogous semiconductor blend characterizations, thereby enabling a more reliable understanding of process-structure relationships for optimized active layer fabrication.

\section{Supporting Information}

Supporting Information is available from the author.

\section{Data Availability}

In compliance with the regulations for projects funded by the German Research Foundation (DFG), the simulation data used for this article is made publicly accessible (see DOI https://doi.org/10.5281/zenodo.17633571).

\section{Author Contributions}
M. Siber: Investigation, Methodology, Formal analysis, Visualization, Conceptualization, Data curation, Writing - original draft, Writing - review and editing.\\
O. J. J. Ronsin: Conceptualization, Software, Supervision, Project administration, Writing - review and editing.\\
G. L. Frey: Conceptualization, Data curation, Supervision, Project administration, Writing - review and editing.\\
J. Harting: Resources, Supervision, Project administration, Writing - review and editing.


\section{Conflicts of Interest} 
There are no conflicts to declare.


\section{Acknowledgements}
The authors acknowledge financial support by the German Research Foundation (DFG, Project HA 4382/14-1), the European Commission (H2020 Program, Project 101008701/EMERGE), the Helmholtz Association (SolarTAP Innovation Platform), and the Collaborative Research Center ChemPrint (CRC1719, Project No. 538767711). 


\printbibliography[heading=bibintoc, title={References}]

\end{document}